\documentclass[a4paper]{article}

\usepackage{icrc2013}
\usepackage[english]{babel}
\usepackage[numbers]{natbib}
\usepackage{upgreek}
\usepackage{amssymb,amsmath}

\newcommand{\F}{\textit{Fermi}}
\newcommand{\g}{$\gamma$}
\newcommand{\xco}{$X_\mathrm{CO}$}
\newcommand{\hi}{$\mathrm{H\,\scriptstyle{I}}$}

\newcommand{\hd}{$\mathrm{H}_2$}
\newcommand{\co}{$\mathrm{CO}$}

\newcommand{\apj}{ApJ }
\newcommand{\apjl}{ApJL }
\newcommand{\apjs}{ApJS }
\newcommand{\mnras}{MNRAS }
\newcommand{\araa}{ARAA }
\newcommand{\nat}{Nature }
\newcommand{\aap}{A\&A }
\newcommand{\pasj}{PASJ }

\newcommand{\ssr}{SSRv }
\newcommand{\prd}{PRD }
\newcommand{\jcap}{JCAP }
\newcommand{\nar}{NewAR }

\title{A tale of cosmic rays narrated in \g~rays by \F}

\shorttitle{A tale of cosmic rays narrated in \g~rays by \F}

\authors{
Luigi Tibaldo$^{1}$
for the \F~LAT collaboration.
}

\afiliations{
$^1$ Kavli Institute for Particle Astrophysics and Cosmology, SLAC National
Accelerator Laboratory
}

\email{ltibaldo@slac.stanford.edu}

\abstract{Because cosmic rays are charged particles scrambled by magnetic
fields, combining direct measurements with other observations is crucial to
understanding their origin and propagation. As energetic particles traverse
matter and electromagnetic fields, they leave marks in the form of neutral
interaction products. Among those, \g~rays trace interactions of nuclei that
inelastically collide with interstellar gas, as well as of leptons that undergo
Bremsstrahlung and inverse-Compton scattering. Data collected by the \F~Large
Area Telescope (LAT) are therefore telling us the story of cosmic rays along
their journey from sources through their home galaxies. Supernova remnants
emerge as a notable \g-ray source population, and older remnants interacting
with
interstellar matter finally show strong evidence of the presence of accelerated
nuclei. Yet the maximum energy attained by shock accelerators is poorly
constrained by observations. Cygnus X, a massive star-forming region established
by the LAT as housing cosmic-ray sources, provides a test case to study the
impact of wind-driven turbulence on the early propagation. Interstellar emission
resulting from the large-scale propagation of cosmic rays in the Milky Way is
revealed in unprecedented detail that challenges some of the simple assumptions
used for the modeling. Moreover, the cosmic-ray induced \g-ray luminosities
of galaxies scale quasi-linearly with their massive-star formation rates: the
overall normalization of that relation below the calorimetric limit suggests
that for most systems a substantial fraction of energy in cosmic rays escapes
into the intergalactic medium. The nuclear production models and the
distribution of target gas and radiation fields, not determined precisely enough
yet, are key to exploiting the full potential of \g-ray data. Nevertheless,
data being collected by \F\ and complementary multi-wavelength/multi-messenger
observations are bringing us ever closer to solving the cosmic-ray mystery.}

\keywords{cosmic rays, diffuse
emission, galaxies, gamma rays: observations, gamma rays:
production, interstellar
medium, Milky Way, superbubbles, supernova remnants.}

\begin{document}
\maketitle

\section{Introduction: on the connection between cosmic rays and \g~rays}

Cosmic rays (CRs), high-energy particles filling the
interstellar space in
galaxies, are a product of the most energetic processes in the
Universe, and also a fundamental constituent of the interstellar medium (ISM)
\citep{ferriere2001}. Despite more than one century of investigation in
this field, we are still struggling to understand which processes accelerate
particles over more than ten decades in energy, which mechanisms rule their
propagation, how they interact with the ISM and influence its evolution, and if
they carry any traces of still unknown physical phenomena such as dark matter
annihilation or decay.

The vast majority of CRs are charged particles, mostly nuclei. Disordered
magnetic
fields scramble their trajectories, so that direct measurements of CRs in the
proximity of the Earth cannot directly identify their production sites.
However, interactions of charged particles with gas and electromagnetic fields
produce neutral secondaries that we can detect and use to trace them from the
point of interaction.

Among neutral secondaries, \g~rays play a crucial role because they trace the
most energetic interactions and are relatively easy to detect, as opposed,
e.g., to neutrinos. The most important processes to produce \g~rays are: 1)
inelastic collisions between high-energy nucleons and gas nuclei that yield in
the final states \g~rays mostly due to the production and decay of neutral
mesons (notably $\pi^0$), 2) Bremsstrahlung of high-energy electrons in the
Coulomb fields of gas nuclei, 3) inverse-Compton (IC) scattering of low-energy
photons (cosmic microwave background, thermal emission from dust, stellar
radiation) by high-energy electrons. 

The \g-ray sky at energies between 20~MeV and hundreds of GeV has been surveyed
with unprecedented sensitivity and angular resolution over the past five years
by the Large Area Telescope (LAT) aboard the \textit{Fermi Gamma-ray Space
Telescope} \citep{latperf2012,atwood2009}. In this paper I will review how LAT
observations are complementing other multi-wavelength and multi-messenger data
to tell us the story of CRs along their journey through the galaxies they
originate from. I will focus
on CRs with energies up to hundreds of TeV that are traced by \g~rays detected
by the LAT. I will summarize the main lessons learned from LAT
observations and challenges for the coming years concerning CR acceleration in
supernova remnants (SNRs), the early phases of CR life in massive-star forming
regions, and the large-scale propagation of CRs in the Milky Way as well as in
external galaxies.

\section{Cosmic-ray acceleration in supernova remnants}

SNRs are strongly advocated to be the most likely sources of CRs at least up to
the
\textit{knee} in their overall spectrum at $\sim 3 \times 10^{15}$~eV
\citep[e.g.,][]{drury2012}. This was
initially motivated by the fact that the observed rate of supernovae in the
Milky Way of $\sim 1/50$~yr$^{-1}$ and their characteristic energy  of
$10^{44}$~J yields, on the assumption of 10\% conversion efficiency into
accelerated
particles, a power of a few $10^{33}$~W, i.e. the CR injection power
inferred from the local CR density of $10^{-13}$~J m$^{-3}$ and the propagation
volume of $10^{62}$~m$^3$ and escape time of $10^8$~yr necessary to explain
the elemental/isotopic abundances in the local CRs \citep{ginzburg1964}. 

This fact turned into the so-called \textit{SNR paradigm} thanks to the
development of a comprehensive theory of nonlinear diffusive shock
acceleration (NDSA), that explains quite successfully how SNRs can accelerate
particles that
subsequently undergo diffusive propagation in the Milky Way to reproduce the
observed CR phenomenology \citep[e.g.,][]{blasi2013}.

Multiwavelength spectra of SNRs undoubtedly show the presence of accelerated
electrons. The acceleration of nuclei was more elusive. It was
unambiguously demonstrated
only very recently for two middle-aged SNRs interacting with molecular
clouds, W44 and IC~443, thanks to the detection of the characteristic
\g-ray spectral signature from $\pi^0$ decay by AGILE \citep{giuliani2011} and
the LAT \citep{LAT2013pidec}. 

However, a question that remains to be answered is the maximum particle energy
that SNRs
can provide. This is important to connect NDSA theory with direct CR
measurement,
since
the CR spectrum does not show any sizable features up to the \textit{knee}.
Hence, we expect a single acceleration mechanism and source class to
dominate. The \g-ray spectrum of the Tycho SNR measured by the LAT
\citep{giordano2012Tycho}, in addition to other multiwavelength
observations, convincingly points to the presence of accelerated nuclei up to
energies of $\sim 0.5\times 10^{15}$~eV \citep{morlino2012}. Indeed, NDSA theory
predicts SNRs to be able to produce the
bulk of CRs up to the knee \citep[e.g.,][]{blasi2012}. This prediction needs to
be further investigated from the observational point of view using LAT data
and data from TeV \g-ray telescopes to pinpoint PeV nuclei in or close to SNRs,
and constrain the poorly understood process of escape from the shock
\citep[e.g.,][]{gabici2009}.

\begin{figure}[tbhp]
  \centering
  \includegraphics[width=0.48\textwidth]{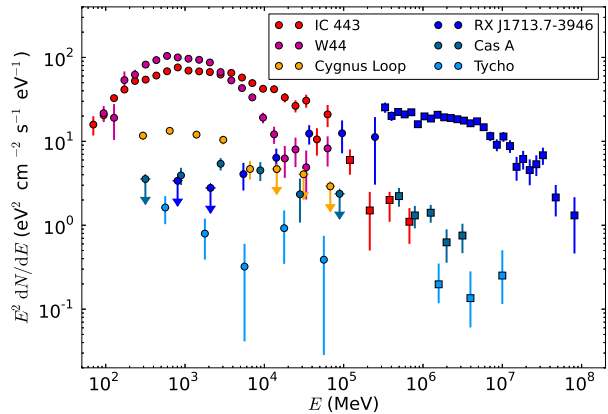}
  \caption{Examples of broad-band \g-ray spectral energy distributions for some
SNRs. Circle points are derived from LAT data, square points from IACT data
(from H.E.S.S., VERITAS, or MAGIC).
The SNRs are: IC~443 \citep{LAT2013pidec,albert2007ic443}, W44
\citep{LAT2013pidec}, the Cygnus Loop \citep{katagiri2011}, RX~J1713.7$-$3946
\citep{abdo2011RXJ1713,aharonian2007rxj1713}, Cas A
\citep{abdo2010casa,humensky2008}, and Tycho
\citep{giordano2012Tycho,acciari2010casa}.}
  \label{SNRspectra}
 \end{figure}

Another major achievement by \F\ is the
possibility to study SNRs as a \g-ray source population
thanks to the detection of a large number of these
objects. Thanks to its superior angular resolution, the
LAT is also the first instrument in the GeV band able to clearly show in some
objects the association between \g-ray emission (i.e., accelerated
particles) and the shock region \citep[e.g.,][]{katagiri2011}.
Figure~\ref{SNRspectra} shows some examples of broad-band \g-ray spectra of SNRs
obtained by combining results from the LAT and data from various imaging
atmospheric
Cerenkov
telescopes (IACTs). 
 
Middle-aged ($\gtrsim 10^4$~yr) SNRs interacting with dense interstellar
material (e.g., IC~443, W44)
are the brightest SNR class in the LAT energy range, while same-age objects
interacting with lower-density matter have lower \g-ray luminosities
(e.g., the Cygnus Loop). Middle-aged SNRs show curved \g-ray spectra,
indicative of a break or cutoff in the underlying particle spectrum at a few
GeV, and the
multiwavelength modeling favors the origin of the bulk of \g-ray
emission from interactions of accelerated nuclei. The origin of the spectral
curvature is not clear yet and may be attributed to the aging of the shockwave
and evolution of properties of escaping particles \citep[e.g.,][]{gabici2009},
or to interactions of the shockwave with the surrounding neutral medium
\citep[e.g.,][]{uchiyama2010,malkov2011}.

Younger SNRs (with ages of centuries to a few $10^3$~yr),  still expanding
in relatively low-density environments, show \g-ray spectra
extending as power laws to TeV energies (e.g., Cas~A, Tycho, RX~J1713.7$-$3946),
with weak evidence for harder spectra relative to older objects
\citep{dermer2013}. This may be tentatively explained by the prediction of NDSA
theory that the maximum acceleration efficiency occurs toward the end of the
initial free-expansion phase (lasting for $\sim 10^3$~years). Some of these
young SNRs are more easily described
by models in which \g-ray emission is predominantly produced by IC interactions
of energetic electrons (e.g., RX~J1713.7$-$3946).

So far we are relying on a limited sample of
cases well-studied individually. Progressing in understanding SNRs as a
CR-source class and
disentangling evolutionary and environmental effects affecting their
\g-ray emission properties will require a comprehensive study
of SNRs as a \g-ray source population. This is the objective
of the forthcoming first LAT SNR Catalog \citep{icrcjack,icrcterri}.

Preliminary results yield the identification of twelve \g-ray
emitting SNRs, and the detection of $>40$ SNR candidates. The aggregate
properties of \g-ray SNRs already begin to challenge some of the simple
assumptions used for the multiwavelength modeling. For instance, while for young
SNRs there is a
correlation between \g-ray and radio spectral indices consistent with
expectations for \g-ray and radio emission produced
by particle populations with the same spectra, for older SNRs interacting with
interstellar material there is a different correlation pointing to two emitting
particle populations with different properties.

\section{Cosmic rays in massive-star forming regions}

The isotopic abundances in CRs show deviations from those in the solar system,
e.g., a ratio $^{22}$Ne/$^{20}$Ne larger by a factor of 5 than in the solar
system, and similar deviations for
trans-iron elements. This points to $\sim 20\%$ of CRs being accelerated in
regions of massive-star formation
\citep[e.g.,][]{binns2007,icrcbinns}. Moreover, $\sim 80\%$ of supernovae are
produced by
the collapse of cores of massive stars, thus often in massive-star clusters, as
opposed to thermonuclear explosions in isolated binary systems.

These indications gave credence to the hypothesis that some of the CRs may be
accelerated in massive-star forming regions by the repeated action of SNR and
stellar-wind shockwaves in \textit{superbubbles} \citep[e.g.,][]{bykov1992}.
Regardless of the acceleration mechanism, if CRs are produced in massive-star
forming regions their early propagation may be significantly influenced by the
turbulent environment characterized by supersonic winds, ionization fronts, and
enhanced radiation fields.

\F\ observations enabled us to image for the first time the early phases of CR
life in the massive-star forming region of Cygnus X \citep{LATcocoon2011}.
Cygnus X, located at
1.5~kpc from the solar system in the Local Arm of the Milky Way, hosts more than
100 O~stars
and several hundred B~stars grouped in numerous clusters formed out of a
reservoir of millions of solar masses of gas.

The LAT revealed an extended excess of \g-ray emission with a
hard
spectrum over the diffuse emission produced by CRs with a spectrum compatible
with that of quiet local interstellar clouds near the solar system
\citep{LATCygISM2012}. The morphology of this excess \g-ray emission closely
follows that of the 50-pc interstellar cavities carved in the Cygnus complex by
the activity of the stellar clusters and bounded by the photon-dominated
regions visible at 8~$\upmu$m \citep{LATcocoon2011}.

This strongly suggests an interstellar origin for the excess, that, however,
cannot be explained under the assumption that it is produced by interactions of
CRs with a spectrum similar to that near the solar system,
neither via nucleon-nucleon interactions (the measured spectrum is too hard)
nor via IC scattering even in the enhanced radiation field in Cygnus X (the
excess is too intense by a factor $\sim 50$ and too hard). The hard spectrum
points to CRs that recently underwent acceleration processes. Therefore, we
concluded that the Cygnus~X cavities form a cocoon of young CRs
\citep{LATcocoon2011}. On the other hand, the CR population averaged over the
whole Cygnus complex on a scale of $\sim 400$~pc, as traced by \F\ measurements,
is consistent with that in the
local interstellar space \citep{LATCygISM2012} (see Figure~\ref{emcomp}),
suggesting an efficient confinement of the particles in the cocoon.

The observations were not sufficient to identify the source of the
freshly-accelerated particles. They may have been accelerated by a SNR, notably
by
G78.2+2.1, that lies in the same direction as the cocoon, even if their mutual
relationship is not clear. Alternatively, they may have been accelerated in a
distributed process by the multiple stellar wind shocks that are energetic
enough and could effectively confine CRs in the region over timescales of $\sim
10^5$~years.

More data to look for possible spectral variations across the cocoon, and
theoretical advances in the modeling of particle transport in superbubble-like
environments are required in order to achieve a deeper understanding of this
phenomenon. However, the results so far confirm the long-standing hypothesis
that massive-star forming regions shelter CR acceleration, and provide a test
case to  advance understanding of CR propagation in presence of wind-driven
turbulence.

Milagro detected potentially extended $>$~TeV emission from Cygnus~X
\citep{abdo2007,milagro2012cyg}, and ARGO extended the measurement down to
$\sim
600$~GeV \citep{icrcargo}. The angular resolution of these two telescopes
was not sufficient to
establish whether there is a more extended diffuse contribution in addition to
emission
from the $0.2^\circ$ source TeV~J2032$+$4130, but they
both measured a flux significantly higher than what IACTs reported for the
latter \citep{albert20082032}. Therefore, this excess emission may be the TeV
counterpart to the \F\ cocoon. The upcoming survey of the northern sky by  HAWC
\citep{goodman2013}, thanks to its improved angular resolution, may shed light
on
the nature of TeV emission from Cygnus X. If a TeV counterpart to the \F\
cocoon is detected, this will help us to locate the particle acceleration
site(s), disentangle whether the accelerated particles are electrons or nuclei,
and constrain their maximum energies.

Other large massive-star clusters exist in the Milky Way and in nearby
galaxies like the Large Magellanic Cloud. GeV and TeV emission was detected
toward a few of
them \citep[e.g.,][]{aharonian2006diff,aharonian2007westerlund2,murphy201230Dor,
abramowski2012westerlund1}. Over the coming years studies of diffuse emission
from massive star forming regions by \F\ and present and future TeV telescopes,
such as HAWC \citep{goodman2013} and the Cerenkov Telescope Array (CTA)
\citep{actis2011}, will be key to understanding the acceleration and early
propagation of CRs in
such a turbulent environment.

\section{Large-scale propagation of cosmic rays in the Milky Way}

The history of large-scale CR propagation in the Milky Way is encoded in
their composition that we can measure directly in the proximity of the solar
system. Ratios of unstable to stable isotopes, such as
$^{10}$Be/$^{9}$Be, serve as radioactive clocks that constrain the residency
time
of CRs in the Milky Way to  $\sim 10^7$~years. Ratios such as B/C show an
overabundance of secondaries produced in inelastic nuclear interactions over
primaries with respect to typical abundances in the solar system, pointing to
interactions of CRs with a few g~cm$^{-2}$ of traversed interstellar material.
These data are interpreted by assuming that CRs propagate in a halo surrounding
the Milky Way occasionally interacting with the denser material in its disk
\citep[e.g.,][]{strong2007}.

It is widely accepted that large-scale CR propagation in the ISM is dominated by
diffusion on disordered magnetic fields, with the possible inclusion of
convection. Over the past years this problem has been treated preferentially
with numerical codes based on simplified yet realistic models of the Milky Way,
such as GALPROP \citep{moskalenko1998,strong2007,icrcmoska}. However, the inputs
to
such
models, for instance the exact value and dependence on particle rigidity of the
diffusion coefficient or the size of the propagation halo, are highly uncertain
\citep[e.g.,][]{strong2007,maurin2010}. Interstellar \g-ray emission produced by
CRs during their propagation can be used to characterize the properties of CRs
(densities, spectra) throughout the Milky Way, and therefore constrain their
origin and transport.

\begin{figure*}[tbhp]
  \centering
  \includegraphics[width=0.75\textwidth]{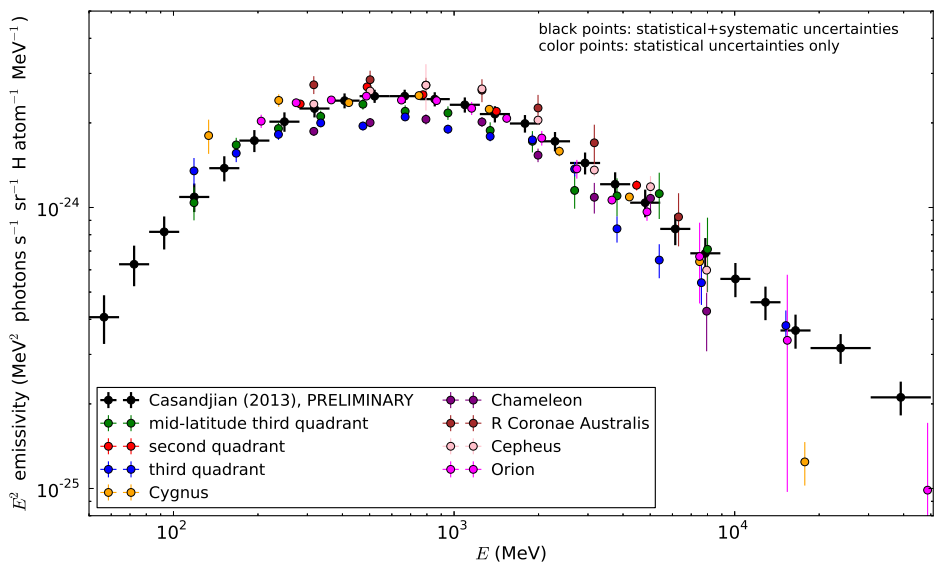}
\caption{Gamma-ray emissivity per H atom in the local interstellar space as
measured for the whole sky at intermediate Galactic latitudes \citep{icrcjm}, a
mid-latitude region in the third Galactic quadrant
\citep{abdo2009lochiemiss}, nearby clouds in the Galactic plane in the second
\citep{abdo2010cascep} and third \citep{ackermann20113quad} Galactic quadrants,
the Cygnus complex \citep{LATCygISM2012}, the R Coronae Australis
and Cepheus clouds \citep{ackermann2012locclouds}, the Chameleon cloud
\citep{ackermann2012loccloudserratum}, and the Orion complex
\citep{LAT2012orion}.
}\label{emcomp}
\end{figure*}

The interstellar space within $\sim 1$~kpc of the solar system is a region
uniquely suitable to study interactions of CRs with well-resolved gas
complexes in the Gould Belt and the local arm. Figure~\ref{emcomp} shows the
\g-ray emissivity, i.e., the \g-ray emission rate per H atom, derived by
comparing LAT intensity maps with maps of the multiwavelength tracers of the
ISM, including the column densities of atomic hydrogen derived from
observations of its 21-cm line \citep[see, e.g.,][]{abdo2010cascep}.

The emissivity traces the densities of CRs and encodes information about their
spectrum. Taking into account the effect of solar modulation, the emissivity of
local interstellar gas is consistent with predictions based on the CR spectra
directly measured near
the Earth, and show that CR densities are uniform within $\lesssim 30\%$ in the
proximity of the solar system\footnote{The emissivity spectrum of the Chameleon
cloud in \citep{ackermann2012locclouds} shows a larger deviation, that,
however, was found to be due to analysis issues. An \emph{erratum} has been
published \citep{ackermann2012loccloudserratum}.}. In fact, the local emissivity
can be
used to
infer the local interstellar spectrum of CRs \citep{icrcjm,icrcchuck} and serve
as a
``Rosetta stone'' to interpret
\g-ray measurements in terms of the underlying CR properties.

For protons the interstellar spectrum can be determined independently from
direct measurements, and used to constrain solar
modulation in conjunction with the first direct measurements at the border of
the heliosphere that are now becoming available \citep{stone2013}. Preliminary
results \citep{icrcchuck} indicate good spectral agreement with
direct measurements in the heliosphere above 10~GeV, with a modest $\lesssim
30\%$ excess in absolute fluxes that may have various origins, including
uncertainties
in the cross sections and the gas tracers, as discussed later, and/or hidden
systematic errors in the direct measurements themselves. This work shows
evidence for
deviations from a simple power law in particle momentum at energies of a few
GeV,
where direct measurements are more complicated to interpret due to solar
modulation, confirming previous clues from elemental abundances
\citep[e.g.,][]{strong2007}.

The large-scale properties of diffuse \g-ray emission from the Milky Way were
recently compared to a suite of GALPROP models \citep{LATdiffpapII}, showing an
approximate agreement
between LAT data and model predictions within $\lesssim 30\%$ over the whole
sky. From this study, a range of source and propagation parameters consistent
with
data was established, but the level of degeneracy is high. In spite of the
overall good agreement, deviations with respect to the models
appear both on intermediate scales due to local peculiarities, such as the
Cygnus~X cocoon where freshly-accelerated particles are injected,  and
on
large scales, highlighting limitations of the current models.

The most remarkable example is the ``\F~bubbles'' \citep{su2010}, giant lobed
structures apparently emanating from the Galactic center filled by energetic
particles emitting in \g~rays. The bubbles are probably powered by activity
in or near the central region of the Milky Way, such as due to a past active
state of the central black hole or to enhanced massive star formation.
Substantial uncertainties in the morphological and spectral properties of the
bubbles due to foreground emission from the Milky Way need to be taken into
account in characterizing the properties of the underlying particle population
\citep{icrcanna}. 

\begin{figure*}[tbhp]
  \centering
  \includegraphics[width=0.49\textwidth]{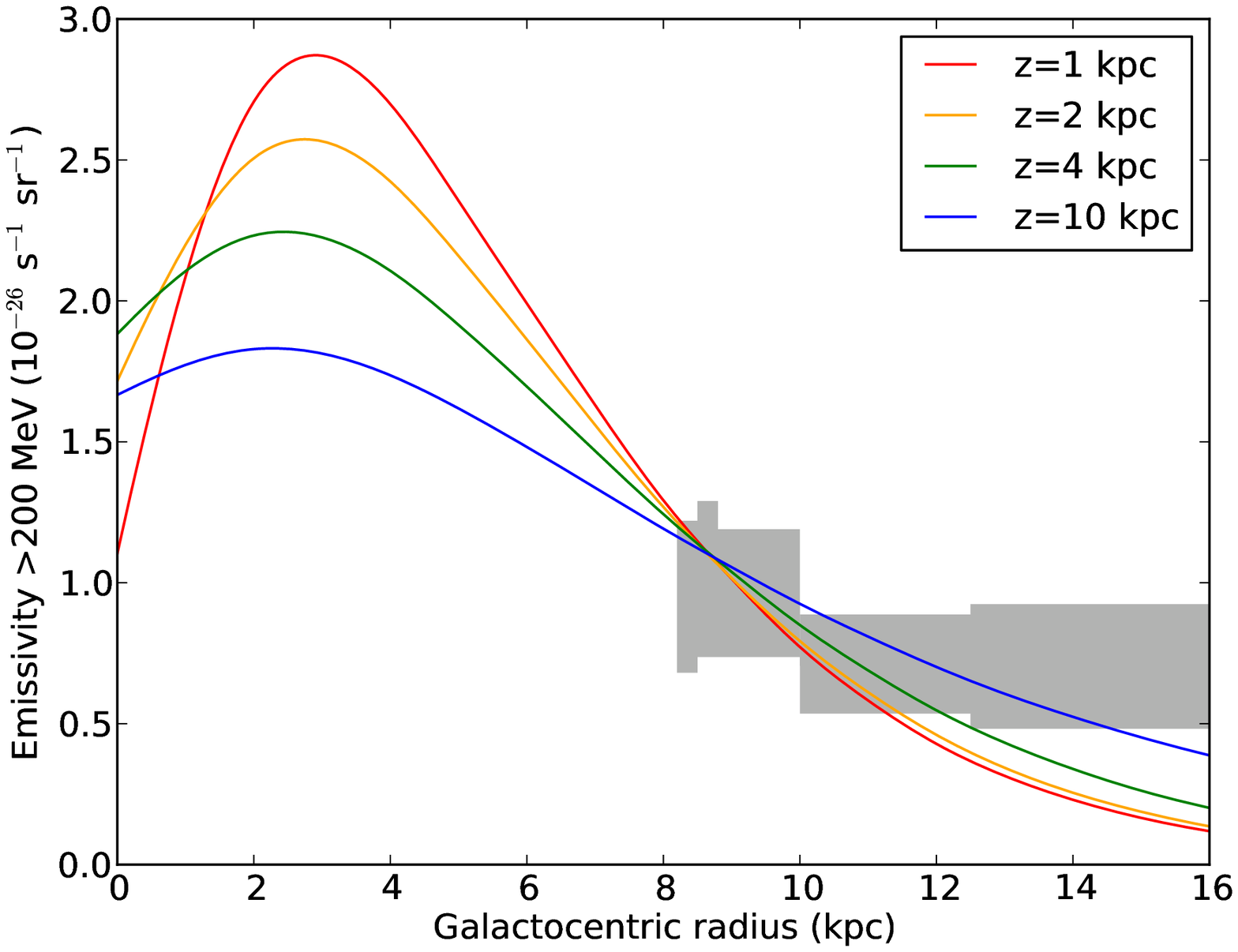}
  \includegraphics[width=0.49\textwidth]{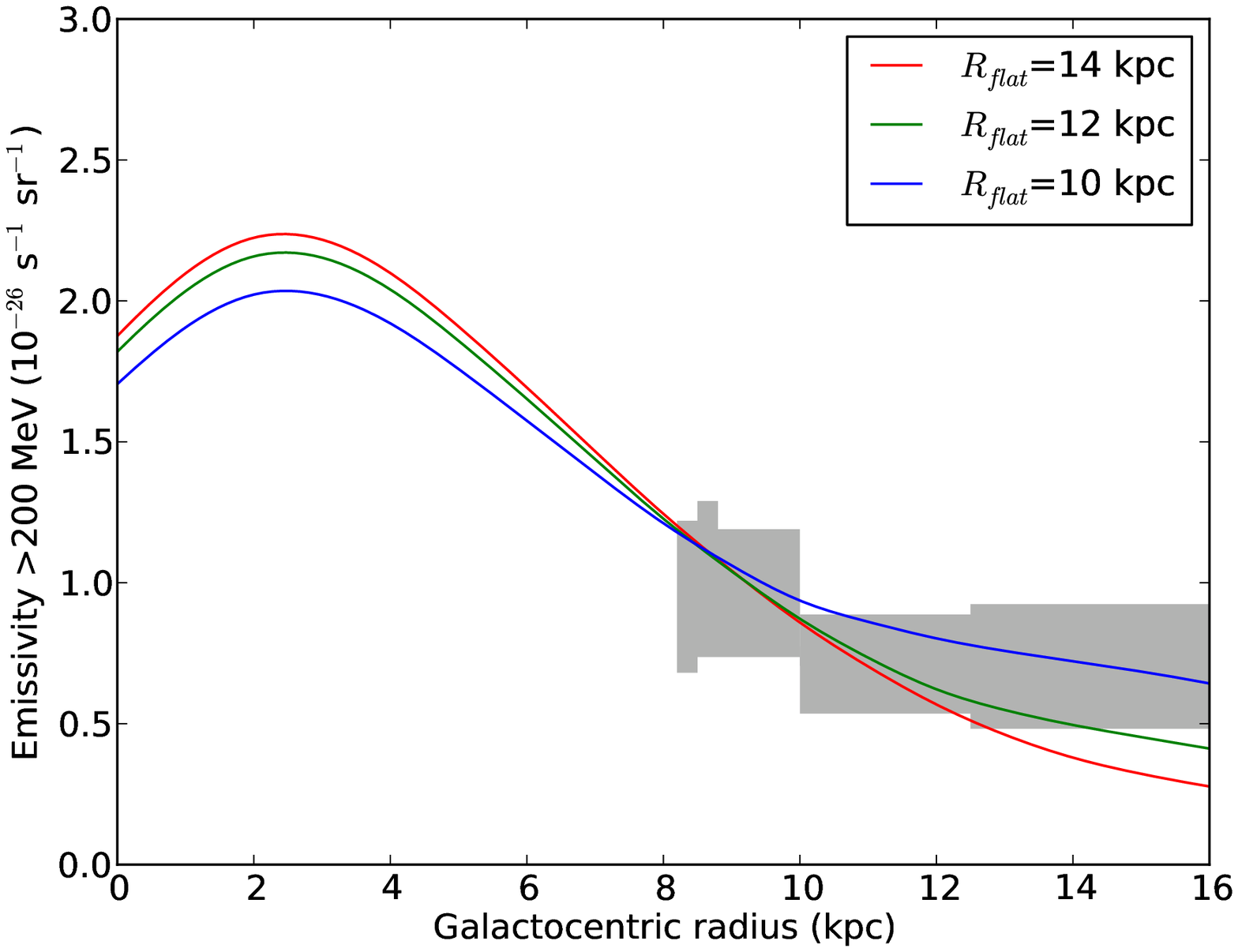}
\caption{The gray band shows the envelope of the measurements of the \g-ray
emissivity at energies $>200$~MeV as a function of Galactocentric
radius from \citep{LATCygISM2012,abdo2010cascep,ackermann20113quad}. The curves
show model predictions from \citep{ackermann20113quad} based on different
values of the height of the CR propagation halo $z$ (left, assuming a CR
source distribution from SNR observations) and on the assumption that the CR
source
density is constant as a function of radius beyond $R_{flat}$ (right, assuming
$z=4$~kpc).}
\label{emprof}
\end{figure*}

Another remarkable example is given by an excess in emission from gas observed
toward the outer region of the Milky Way
\citep[][]{abdo2010cascep,ackermann20113quad}, where the density of putative CR
sources, SNRs or regions of massive-star formation, is greatly reduced.
Figure~\ref{emprof} shows the profile of \g-ray emissivity as a function of
Galactocentric radius. We used the Doppler shift or radio lines
tracing interstellar gas to
separate different cloud complexes
along the line of sight, and, subsequently, to evaluate their \g-ray emissivity
thanks to the morphological resemblance to the radio maps
\citep[e.g.,][]{abdo2010cascep}.

Figure~\ref{emprof} shows two possible solutions to the \emph{gradient
problem} in the outer Galaxy, i.e., the flat emissivity profile: a
thick CR propagation halo with a height of $\sim 10$~kpc (at the upper limit of
the
range allowed by the $^{10}$Be/$^{9}$Be ratio), or a flat CR source density in
the outer Galaxy, at odds with multiwavelength observations of the putative
sources. This issue may point again to the
inadequacies of simplified propagation models to reproduce the detailed
structures of the \g-ray sky, e.g., in this case, due
to spatial variations of the diffusion coefficient possibly related to the
dynamical coupling between CRs and interstellar turbulence
\citep{breitschwerdt2002,evoli2012}.

\section{Cosmic rays in external galaxies and beyond}
Interstellar \g-ray emission is produced in external galaxies in the same
way as in the Milky Way. Observations of external galaxies are very useful
because they are free from the complications of viewing the Milky Way from
inside and are well suited to study the properties of the galaxies' aggregate
emission. On the other hand, the limited angular resolution of the \g-ray
instruments makes it difficult to separate interstellar emission from individual
objects within galaxies, and especially from active galactic nuclei that are the
largest known class of \g-ray sources.

Historically, observations of nearby galaxies by EGRET \citep{lin1996}
demonstrated from the observational point of view that CRs below the \emph{knee}
are galactic in origin, since galaxies' \g-ray fluxes are not consistent with
particle densities constant over the local group, as confirmed later by \F\
\citep{abdo2010lmc,abdo2010smc}.

The LAT brought a new quality to the subject, first of all making
spatially resolved studies of nearby galaxies
possible
thanks to the improved angular resolution \citep{abdo2010lmc,abdo2010smc}.
Studies of the Large
Magellanic Cloud based on LAT data \citep{abdo2010lmc,murphy201230Dor} showed
how diffuse \g-ray emission intensities peak toward 30~Doradus, a conspicuous
region of massive-star formation. Moreover, the \g-ray intensities from the
Large Magellanic Cloud do not correlate well on a global scale with its neutral
gas densities \citep{abdo2010lmc}.

The LAT, with the detection of a few nearby galaxies, and more distant starbust
galaxies, also enabled us to perform the first population studies
in \g~rays \citep{LATSFGalaxies}, and look for correlations with their
luminosities at other wavelengths. LAT detections and upper limits on the
fluxes from a sample of galaxies selected
based on their star-formation rates \citep{LATSFGalaxies} revealed a
quasi-linear scaling between \g-ray luminosities and either radio or infrared
(IR) luminosity in the 8--1000~$\upmu$m band. The correlation is
robust against the inclusion or not in the analysis of galaxies with known
active nuclei. 

This resembles the well-known IR-radio correlation
\citep[e.g.,][]{jarvis2010}.The latter is interpreted based on the idea
that galaxies are electromagnetic \emph{calorimeters}, which means that all the
energy injected in the form of leptonic CRs is dissipated within galaxies and
re-emitted in form of electromagnetic radiation, including radio synchrotron
emission. Therefore, if CR acceleration is related to massive stars and/or
the aftermaths of their final explosions, SNRs, the radio
luminosity is expected to be proportional to IR luminosity, a proxy for
star-formation rate
due to dust heating by stellar radiation.

On the other hand, the quasi-linear relation between \g-ray and IR luminosity
lies below
the calorimetric limit derived assuming that all the energy injected in the
form of hadronic CRs\footnote{The estimate requires assumptions on the
relation between IR/radio luminosity and star-formation rate, and also
assumes the canonical supernova energy of $10^{44}$~J converted with $10\%$
efficiency into accelerated particles. See \citep{LATSFGalaxies} for details.}
is dissipated within galaxies and re-emitted in the \g-ray
domain. This possibly indicates that the final fate of most of CR nuclei is
escaping from galaxies into the intergalactic medium, as formerly inferred from
CR propagation models for the Milky Way \citep{strong2010}. Observational
evidence based on 4 yr of LAT data for the presence of CR nuclei in the
intergalactic medium of galaxy clusters, injected by either member galaxies or
intergalactic processes, is still very weak \citep{LAT2013clusters}.

Starbust galaxies, with high rates of massive-star formation, are the only ones
found
to be close to the calorimetric limit in \g~rays. In contrast to more quiet
local
galaxies, they
also have \g-ray spectra harder than the Milky Way, extending in some cases
to TeV energies \citep[e.g.,][]{veritas2009m82}. These facts suggest that in
starbust galaxies energy losses of CR nuclei may not be dominated by diffusive
transport, as in the Milky Way, but rather by energy-independent mechanisms.

\section{What else do we need to get the most out of \g-ray observations?}

The LAT brought high-energy \g-ray astronomy from a science still dominated by
limited photon statistics to a point where systematic uncertainties often play
the dominant role. In addition to uncertainties inherent to the \g-ray
measurements, like
those in the instrument effective areas or backgrounds, there are
some that depend from other measurements. I will summarize here the
most relevant ones that we could improve on in the near future in order to
exploit the full potential of \g-ray measurements to learn about CR physics.

Uncertainties can be attributed to the targets for \g-ray production,
both gas and interstellar radiation fields. While substantial progress in the
interpretation of \g-ray observations for
individual objects, such as SNRs, can be often achieved only on a case-by-case
basis, some general properties of ISM tracers may soon be better constrained by
observations and advances in theory/simulations, helping us to better understand
CR
nuclei using \g-ray data.

Interstellar gas is most often traced by combining the 21-cm line of \hi\ to
trace the neutral atomic gas, the 2.6-mm line of \co\ as a surrogate of
neutral molecular gas, and different dust tracers that serve to account for gas
invisible to the two lines (either opaque or self-absorbed \hi\ or CO-dark
\hd). All of these tracers have known limitations. Forthcoming
high-resolution \hi\ data \citep{kalberla2010} and improved dust maps obtained
from the \textit{Planck} survey \citep[e.g.,][]{Planck2011dg} for the whole sky
are going to be of great use for \g-ray astronomers.

A major source of uncertainty in the determination of the CR content of
the Milky Way at large scale currently is the determination of
\hi\ column densities from the 21-cm line, subject to approximations in the
analytic handling of the radiative transfer equation
\citep[e.g.,][]{ackermann20113quad,tibaldosci}.
There is no clear path to improve on this issue, but it may profit from the
combination of the aforementioned survey releases and numerical simulations of
the ISM \citep[e.g.,][]{saury2013} to set the treatment of \hi\ opacity on a
more physical ground than the current assumptions of effective optical
thickness of the neutral medium or of a uniform spin temperature. 

\begin{figure}[tbhp]
  \centering
  \includegraphics[width=0.54\textwidth]{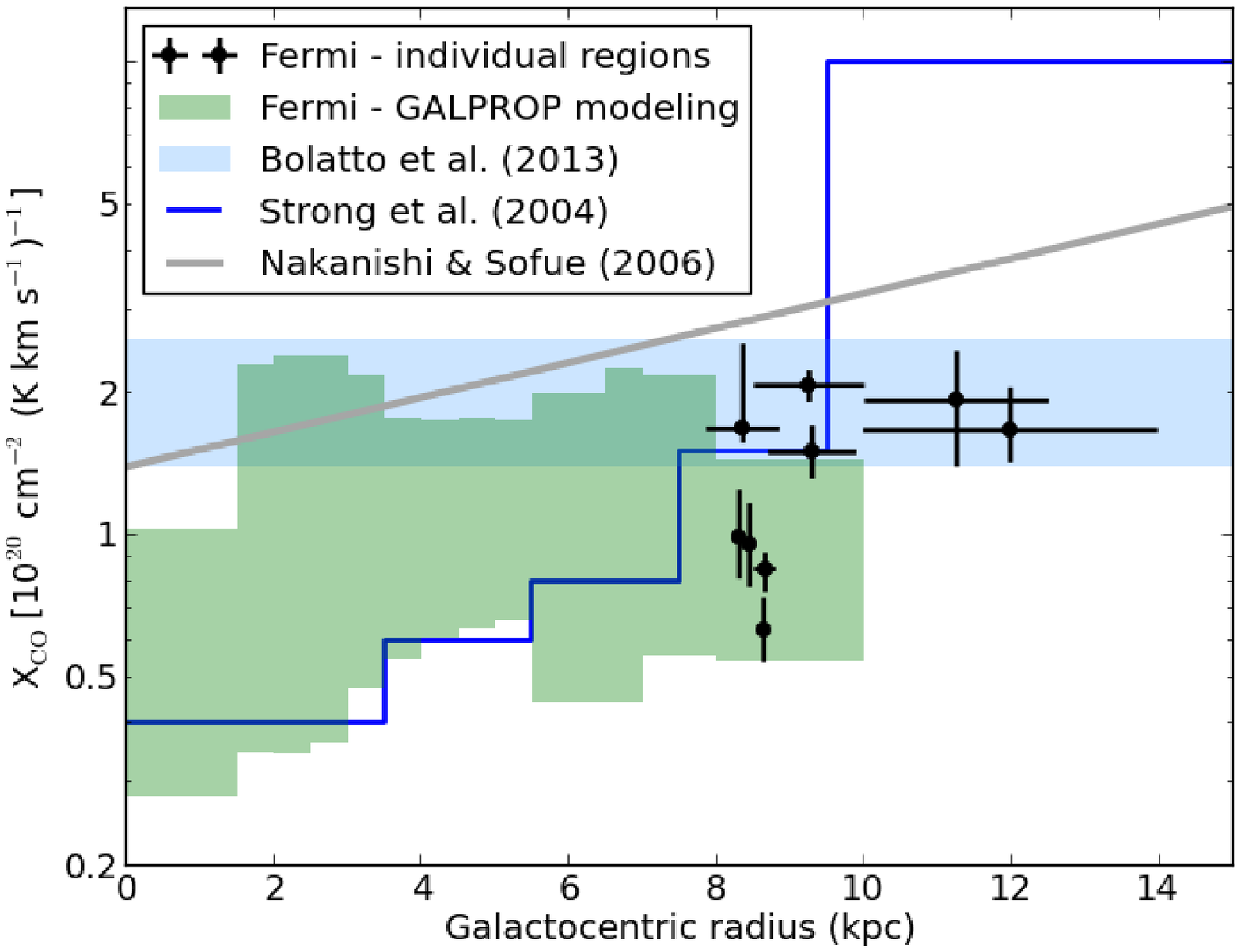}
\caption{\xco\ as a function of Galactocentric radius in the Milky Way. Black
points are derived from LAT observations of selected clouds and complexes
\citep{LATCygISM2012,abdo2010cascep,ackermann20113quad,ackermann2012locclouds}.
The green band shows the range of \xco\ values used to model large-scale
diffuse \g-ray emission as observed by the LAT using GALPROP
\citep{LATdiffpapII}. The light blue band represents the
value for \xco\ in the Milky Way and its uncertainty
recently recommended by \citep{bolatto2013}. The
blue \citep{strong2004grad} and gray \citep{nakanishi2006} lines are two models
not based on \g-ray data used in the literature to study diffuse \g-ray
emission.}
  \label{xcofig}
 \end{figure}

Molecular clouds provide more localized targets for CR interactions, that could
be useful, in principle, to identify intermediate-scale enhancements of
particles.
Unfortunately, the column densities of \hd\ present larger uncertainties, of a
factor of a few on average, since we rely on \co\ as a surrogate tracer.
Figure~\ref{xcofig} shows \xco, the ratio between \hd\ column densities and the
intensities of the CO line, as a function of Galactocentric radius in the
Milky Way, combining results from LAT observations with other values/models from
the literature.
 
 Indeed, \g-ray measurements can be used to derive \xco\ as the emissivity
per CO intensity unit divided by twice the emissivity per H atom for atomic
clouds at approximately the same locations. This assumes that CRs can
penetrate molecular clouds uniformly to their cores and that the variations of
CR densities are small over the scales of interstellar complexes. Note that the
green band in Figure~\ref{xcofig} from \citep{LATdiffpapII}, on the other hand,
represents a range of values used to model LAT data based on some \emph{a
priori} assumptions on the CR distribution in the Milky Way; thus it will be
possible to extract more stringent constraints on \xco\ from LAT data. 
 
 LAT observations are consistent with a constant \xco\ for clouds in the disk
of the Milky Way within a few kpc of the solar system \citep{bolatto2013}.
However, the \xco\ values inferred from
LAT data for well-resolved nearby clouds are lower by a factor $\sim 2$
\citep{abdo2010cascep,ackermann2012locclouds}. The origin of this difference is
still under investigation. There is an indication
from \g~rays that \xco\ drops by about one order of magnitude
in the innermost region of the Milky Way, in agreement with
infrared observations of external galaxies \citep{sandstrom2012}.
Figure~\ref{xcofig} also illustrates how estimates
of \xco\ based on observations at other wavelengths (like those from virial
masses inferred from the 2.6-mm \co\ line, or from infrared emission from dust
as a mass tracer) need to be critically evaluated in relation to \g-ray
observations.

Targets, however, are not the only issue we need to deal with. While the \g-ray 
yield from electromagnetic interactions can be calculated with high precision,
there exist significant uncertainties related to nucleon-nucleon interactions,
that are, in some cases, comparable to uncertainties in the \g-ray measurements
by the LAT \citep{icrcjm,icrcchuck}.

Nuclear production models rely on a set of measurements that are limited in
energy coverage, angular coverage around the interaction point, and
bullet/target elements. Theoretical frameworks bridging those limited sets of
measurements are used to predict \g-ray yields over all energies, directions and
from interactions of all elements in CRs and the ISM. Yet differences between
alternative nuclear production models in some cases turn out to be larger than
the uncertainties intrinsic to the \g-ray measurements
\citep[e.g.,][]{kachelriess2012,dermer2013CRgamma}.

Recent accelerator-based measurements are helping to constrain nuclear
production models at the highest energies accessible \citep{sato2012}. Those
measurements are very helpful to interpret TeV \g-ray data. Nuclear production
models for lower energies of interest to interpret LAT data, however, depend
mostly on accelerator measurements from the '60s and the '70s
\citep[][]{stecker1973,dermer1986b}.
In order to fully exploit the potential of \F\ data to understand CR physics,
nuclear
production models need to be critically evaluated against the existing
accelerator data, and the relative uncertainties need to be taken into account
in \g-ray analyses and propagated to the final results. Accelerator experiments
dedicated to reduce these
uncertainties in the energy range most relevant for \F\ would also be highly
beneficial to the \g-ray community. 

\section{Looking forward: a \g-ray bright future}

Despite all the uncertainties, as I discussed in this article data from the
LAT, together with other multiwavelength and multimessenger observations,
are illuminating several aspects of CR life, and are going to
continue being a treasure trove in coming years.

This will be facilitated by continued accumulation of \g-ray data by the LAT,
especially
in the energy range $>10$~GeV where skymaps are still photon starved, bridging
satellite-based observations to those by current and future ground-based
instruments such as CTA and HAWC. The highest-energy end of the LAT energy
range is interesting both to study CR electrons through IC emission from the
Milky Way that becomes dominant over the emission from gas, and to search for
further evidence of freshly accelerated CRs in selected regions following the
detection in Cygnus~X.

The ongoing revision of the LAT
event-level analysis and subsequent event selection based on the experience
with data gained in the prime phase of the mission \citep{atwood2013P8} is going
to bring a new quality to LAT data, providing increased acceptance for \g-rays,
and a better control of backgrounds and instrumental systematics. The
anticipated improvements in the
angular resolution at low energies will also be key to disentangling sources and
diffuse emission in the Milky Way disk, especially toward massive-star forming
regions and the inner Galaxy. 

This will be most useful in the
energy range $<100$~MeV to bridge LAT observations
with those by Compton telescopes such as COMPTEL \citep{schoenfelder2004} and
potential future missions such as GRIPS
\citep{greiner2012GRIPS}. This will enable us to study in more detail the
spectral roll off of the $\pi^0$ decay component due to nucleon-nucleon
interactions, both in discrete sources such as SNRs and in diffuse emission
from the ISM. Diffuse emission produced by Bremsstrahlung is going to be a
probe of the CR electron population complementary to IC and synchrotron
emission observed in radio to microwaves \citep[e.g.,][]{orlando2013}.

In conclusion, with the many facets of CR acceleration in SNRs, the recent
observations of freshly-accelerated CRs in massive-star forming regions,
interstellar emission from the Milky Way revealed in unprecedented detail, a
growing population of star-forming galaxies detected in \g-rays, and  possibly
new phenomena related to CRs awaiting discovery thanks to the ongoing survey
by \F\ and complementary surveys by the current generation of ground-based
\g-ray telescopes, as well as the forthcoming surveys by HAWC and CTA, the
future of
\g-ray astronomy looks bright and promises to bring us ever closer to
understand the mysteries of CRs.

\vspace*{0.5cm}
\footnotesize{The \F~LAT Collaboration acknowledges support from a number
of agencies and institutes for both development and the operation of the LAT as
well as scientific data analysis. These include NASA and DOE in the United
States, CEA/Irfu and IN2P3/CNRS in France, ASI and INFN in Italy, MEXT, KEK, and
JAXA in Japan, and the K.~A.~Wallenberg Foundation, the Swedish Research Council
and the National Space Board in Sweden. Additional support from INAF in Italy
and CNES in France for science analysis during the operations phase is also
gratefully acknowledged.}


\end{document}